**Draft version 1, 10/2/2023.**

**Note: This paper has not been peer reviewed and is subject to change.**

# Dimensions of Disagreement: Unpacking Divergence and Misalignment in Cognitive Science and Artificial Intelligence

**Kerem Oktar[1], Ilia Sucholutsky[2], Tania Lombrozo[1], and Thomas L. Griffiths[1,2]**

[1]Department of Psychology, Princeton University

[2]Department of Computer Science, Princeton University

**Abstract**

The increasing prevalence of artificial agents creates a correspondingly increasing need to manage disagreements between humans and artificial agents, as well as between artificial agents themselves. Considering this larger space of possible agents exposes an opportunity for furthering our understanding of the nature of disagreement: past studies in psychology have often cast disagreement as two agents forming diverging evaluations of the same object, but disagreement can also arise from differences in how agents represent that object. AI research on human-machine alignment and recent work in computational cognitive science have focused on this latter kind of disagreement, and have developed tools that can be used to quantify the extent of representational overlap between agents. Understanding how divergence and misalignment interact to produce disagreement, and how resolution strategies depend on this interaction, is key to promoting effective collaboration between diverse types of agents.

**Dimensions of Disagreement: Unpacking Divergence and Misalignment in Cognitive Science and Artificial Intelligence**

What is disagreement? It is intuitive to think of it as a divergence of judgment: If Deniz believes that vaccines are safe and Sade does not, then they disagree. This intuitive notion of divergence undergirds much work on disagreement in judgment and decision-making research (e.g., Reeder et al., 2005), as well as political science, social psychology, and epistemology (Carothers & O'Donohue, 2019; Iyengar et al., 2019; Frances & Matheson, 2019). Here, we argue that developments in artificial intelligence and computational cognitive science highlight another dimension of disagreement—representational misalignment—that formalizes ideas with roots in philosophy and developmental psychology (e.g., Carey, 1985; Kuhn, 1962), and that has important implications for conflict resolution and collaboration. Before discussing how JDM and AI research on disagreement can enrich one another, we begin with a historical case study to illustrate these notions and explain why this distinction matters.

**Divergence**

In 1663, Galileo was convicted of heresy by the Roman Catholic Inquisition for his belief that the sun is the center of the universe, as Pope Urban VIII (the voice of God on Earth) maintained that the Earth is the center instead (heliocentrism vs. geocentrism; see Finocchiaro, 2014). Galileo (G) and Urban (U) clearly disagreed. Through a Bayesian lens, whereby beliefs are conceptualized as subjective probability assignments, we can characterize the extent of this disagreement through divergences in their credences about whether the sun is the center of the world (S; e.g., divergence = | $P_G(S) - P_U(S)$ |; see Oktar & Lombrozo, 2023).

Divergence parsimoniously captures the typical way disagreement has been conceptualized and operationalized in psychological research, from disagreement over policy preferences (Reeder et al., 2005) to statistical estimates (Minson et al., 2011) and aesthetic judgments (Cheek et al., 2021), among others. Minson et al. (2011), for instance, operationalize disagreement as the quantitative differences in a dyad's estimates (e.g., about the average income of Israeli families).

**Misalignment**

Divergence by itself fails to capture discrepancies in the algorithms and representations people use to generate judgments. For instance, early helio- and geocentrist models of the universe differed greatly in the astronomical structures they implied, despite making convergent predictions about the apparent positions of planets in the solar system relative to the Earth (see Gearhart, 1985). As a consequence, if asked to generate predictions about planetary observations, both an early helio- and a geocentrist would give similar answers. If assessed via a measure of divergence like the one introduced above, the probabilities they would assign to different events such as a solar eclipse on a particular date would diverge by very little. However, they would reach these beliefs through differing, misaligned representations of the universe, and ultimately offer different explanations for why they believe what they believe (see Figure 1).

**Figure 1**

*Geo- and Heliocentric Models of Planetary Orbits*

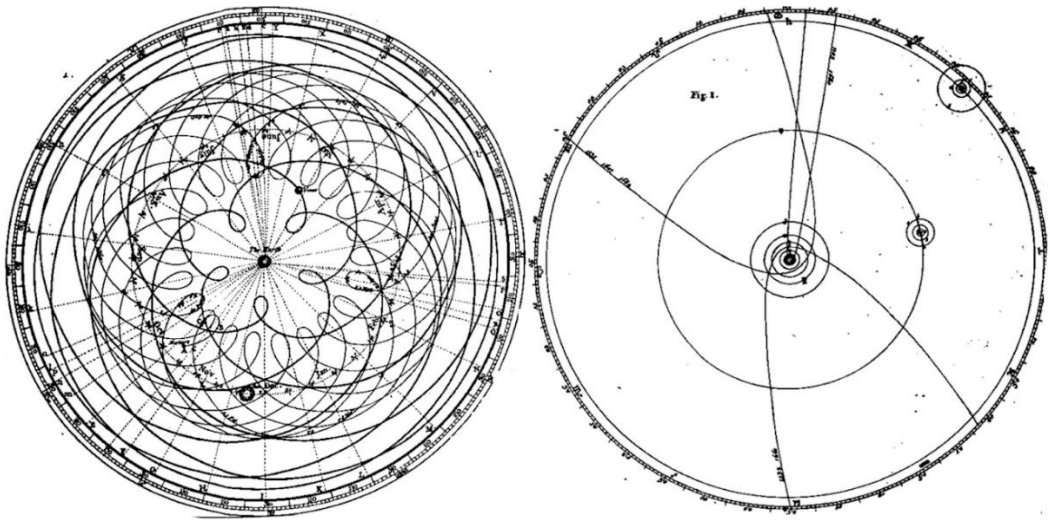

*Note.* Geocentric (left) vs heliocentric (right) models of orbits. Adapted from Ferguson (1756).

The scope of representational misalignment extends far beyond the cosmological. From variation in our experience of basic sensations to our understanding of abstract concepts, diversity in the structure of mental representations has challenged theories of knowledge, communication, and ethics for millennia (often in the form of relativism; Meißner, 2023). Developmental, educational, and cross-cultural psychologists have faced the daunting task of characterizing intuitive theories and mechanisms of change when these theories are not only distinct, but potentially incommensurable (e.g., Carey, 1991; Vosniadou et al., 2008). In the case of sensation, for instance, there can be variation in the phenomenal experience of the same stimulus (e.g., what seems red to me may seem green to you). Such misalignment is philosophically challenging, but practically unproblematic if the relative structure of internal representations remains the same (e.g., we both agree that red is more similar to purple than to green). Thus, it is the *structure* of representations that determines possibilities for communication and collaboration (see Figure 2).[1]

As a result, an intuitive way to define representational misalignment is via a measure of the dissimilarity between the internal representations of a set of stimuli for a given set of agents. For example, neuroscientists measure the correlation of pairwise distances between the two sets of activation vectors produced by two (artificial or biological) neural networks when exposed to the same stimuli (Representational Similarity Analysis – Kriegeskorte et al., 2008). The empirical study of the similarity structure of internal representations has a rich history in cognitive science as it enables inference about internal representations without having direct access to those representations (see Shepard, 1980; Tenenbaum et al., 2011). Classically, pairwise similarity judgments are elicited across a fixed set of stimuli, with items rated as more similar interpreted as being closer to each other in representational space. Recent developments enable even more data-efficient recovery of representations (see Marjieh et al., 2023; Sucholutsky et al., 2023).

---

[1] We can formally frame this mismatch in the following way. If we consider each stimulus as being internally encoded as a vector (e.g., of neural activations), variation in sensation would correspond to differences in the absolute values of these vectors: For instance, your red vector may be equivalent to my green vector. What matters for alignment is whether the relative distances between vectors is preserved across the two representational spaces.

**Figure 2**

*Computing Representational Alignment from Pairwise Similarity*

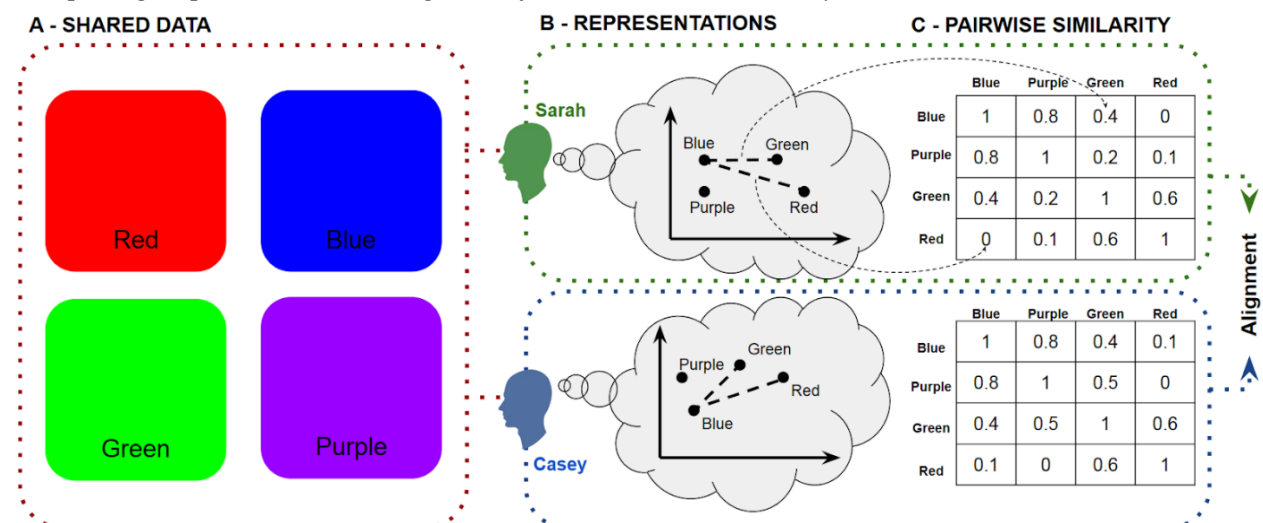

*Note.* Representational alignment as a measure of differences in pairwise similarity judgments. Though Sarah and Casey perceive individual colors differently, their representations are highly aligned, as captured by the correspondence in their pairwise similarity matrices. Adapted from Sucholutsky and Griffiths (2023).

## Why Distinguishing Divergence and Misalignment Matters

Importantly, whether, when, and how disagreements can be resolved depends on the interaction of divergence and misalignment. From a divergence-focused perspective, for instance, a straightforward approach to reducing disagreement is providing disagreeing agents with a common set of data that relate to the issue in question. This is a prevalent strategy for disagreement reduction known as the "deficit model" in science communication (Simis et al., 2016; see also Farrel et al., 2019; Hartman et al., 2022). The intuitive appeal of this strategy has led psychologists to declare that increasing disagreement given the same set of data would lie "in contrast to any normative strategy imaginable for incorporating new evidence relevant to one's beliefs" (Ross & Anderson, 1982, p. 145).

Yet such polarization can be both common and rational in the presence of misalignment. Differences in the set of alternative hypotheses being represented, or the conditional dependencies between hypotheses and data being considered, can lead to divergent conclusions from the same data (Jaynes, 2003, Ch. 5.2; see also Jern, Chang, & Kemp, 2014). In the case of astronomy, for example, the same stellar observations led scholars to polarized conclusions. Whereas geocentric models represented stars as being relatively close to earth, heliocentric models took them to be very far. Thus, the observation that stars remain the same size year-round confirmed both geocentrists' views (if the earth is in the center, stars should be the same size as they are always equidistant), as well as heliocentrists' views (if the stars are very far, they will appear to be the same size since the orbit of the earth is too small to make an observable difference; see Grant, 2007). Of course, data may help resolve disagreement when agents' representations make differing predictions about the same phenomenon (but cf. Gershman, 2019).

**Implications of the Distinction for JDM Research**

As mentioned above, much research in JDM takes a divergence-first approach to disagreement. Yet taking misalignment into account can enrich current lines of inquiry while raising novel questions about the nature of disagreement—and recent research in computational cognitive science has laid the groundwork for these developments. We consider each in turn.

*Enriching Extant Research*

Navigating disagreement is a key component of social intelligence: Negotiating differences in views allows us to learn from others, leveraging their experiences to expand our beliefs beyond what we can observe. We are thus equipped with a suite of socio-cognitive mechanisms highly attuned to others' expertise, social standing, and intentions (Hermann et al., 2007; Mercier & Sperber, 2011). Accordingly, research in JDM has focused on uncovering these mechanisms and characterizing when and how we learn from disagreement, or fail to do so. For example, research on advice-taking has investigated how people weigh their own judgments vs. those of others in estimation tasks (Bonaccio & Dalal, 2006) as well as those of human vs. algorithmic advisors (Glikson & Woolley, 2020). People typically overweight their own judgments in these studies, and many mechanisms have been suggested to account for this egocentric bias, from asymmetric access to reasons (Yaniv & Kleinberger, 2000) to biased sampling (Hütter & Ache, 2016) and motivated reasoning (Kappes et al., 2020). Misalignment offers a distinct and synergistic explanation for biased advice-taking.

When receiving advice, people jointly learn from, and about, their advisors (Bovens & Hartman, 2003). For example, if an advisor provides contradictory advice about similar problems, we might simultaneously use their advice and grow suspicious of their reliability (see Orchinik et al., 2023). Beyond merely estimating reliability, we might also use their estimates to infer their representation of the problem space—in a simple estimation task with one predictor and one outcome, for instance, we could sample the advisor's estimates across possible values of the predictor to infer their representation of the function relating the two variables (grid approximation). Subsequently, learning that an advisor has a complex, misaligned representation could lead to risk-averse discounting of their advice upon error, especially if we take our own representation of the domain to be robust and generalizable. The literature on algorithm aversion shows exactly this pattern: Algorithms are not discounted until they make unexpected and atypical mistakes, after which people quickly lose confidence in them (Dietvorst et al., 2015; cf. Logg et al., 2019). Beyond simple heuristics based on divergence, inferences of misalignment can thus contribute to understanding how people utilize others' advice.

### Raising Novel Questions

Reducing disagreement to divergence simplifies inferences of and from disagreement—and incorporating misalignment raises questions by complicating this analysis. Whereas divergence can be approximated with one sample or communicative act, misalignment is much more difficult to estimate.[2] Minimally, it requires observing systematic divergence across a range of judgments. Maximally, it entails fully simulating the other agent's internal representation of the task. This naturally raises the question of if, when, and how people go through this more informationally—and computationally—intensive inference process, rather than using divergence-based heuristics. An important factor may be the ease of generalization from one's own internal representation to that of the other agent. Generalizing to similar intelligences in well-known domains may be the easiest case since one's own representations can be leveraged to estimate alignment. In the case of disagreeing humans, for instance, this simulative perspective-taking process can co-opt Theory of Mind mechanisms (Frith & Frith, 2005; Goldman, 2006). Similarly, in domains such as intuitive psychology or physics, people's structured representations can be used as informative priors, facilitating alignment inferences (Lake et al., 2017). Disagreements with artificial agents in unfamiliar domains (e.g., protein folding; Atz et al., 2021) pose much tougher challenges for estimating alignment.

---

[2] Relatedly, teachers are fairly accurate at tracking what students know and do not know, but much less accurate at recognizing their alternative understandings and models (Chi et al., 2004).

A related question is how people represent disagreement in groups: In the case of divergence, disagreement across agents can be inferred simply from variance across estimates. Computations of misalignment can also be generalized to multi-agent settings, but the difficulty of computing misalignment in dyads suggests that people may have to rely on increasingly coarse approximations of misalignment, if they take it into account at all. For instance, agents can be initially clustered into subgroups based on their opinions, and their representations can be coarse-grained to be invariant across individuals in each cluster (e.g., our astronomical example caricatured geocentrists and heliocentrists as uniformly believing two distinct models, when in reality there were many competing models that interpolated between the two extremes; Grant, 2007). Ironically, attempting to develop a nuanced understanding of disagreement may backfire in such settings, with caricatured inferences of others' representations contributing to stereotypes.

As for inferences from disagreement, misalignment raises new questions about the striking tendency for individuals to persist in their individual beliefs amid dissent (e.g., roughly 90% do not question their views upon contemplating societal disagreement; Oktar & Lombrozo, 2022). A common path to persistence is subjectivity: If I believe that euthanasia is morally permissible and that moral beliefs are matters of subjective opinion, I may persist in my views amid disagreement by appealing to subjectivity. But what grounds such inferences? One possibility is that subjectivity tracks representational diversity—in domains where there is a lot of variance in how people perceive issues or stimuli (e.g., on abstract notions like morality or love), people may expect disagreement to be incommensurable by default, whereas domains with representational uniformity (e.g., formally defined systems like games, financial markets, or mathematics) may prove more conducive to conciliation.

*Recent Developments*

Work in computational cognitive science has recently shown that misalignment may be much more prevalent than it may intuitively seem. Marti et al. (2023) collected conceptual similarity ratings and feature judgments from a variety of words relating to common animals and politicians, and computed potential clusters of representations of these categories. These analyses reveal substantial and introspectively inaccessible variation in people's word representations: Whereas people erroneously believe that other people share their semantics, their analyses suggest that 10-30 variants of common concepts for even simple categories exist in the population. As noted above, misalignment can interfere with disagreement resolution—especially when it is unperceived. Relatedly, the extent of representational misalignment for word meanings predicts failures of communication across people (Duan & Lupyan, 2023). These studies lay the groundwork for future research on how perceived and actual misalignment combine to influence conflict resolution and collaboration across influential domains.

**Implications of the Distinction for AI Research**

Artificial intelligences can have radically different representations from humans, due to differing goals, as well as constraints on the information and computation available (see Griffiths, 2020). Accordingly, the question of alignment has received much attention in recent academic AI research (e.g., Bommasani et al., 2021; Gabriel, 2020), and has become a focus in industry (with OpenAI recently investing 20% of its compute on a new "superalignment" team; Leike & Sutskever, 2023). This attention is driven in part by the understanding that misaligned representations can precipitate disagreements (Russel, 2019), and in part by the increasing interest in developing collaborative teams of artificial and human agents (Sharma et al., 2023), as representational alignment can increase the rate of convergence when agents need to learn from one another (Sucholutsky & Griffiths, 2023).

Despite broadly increasing attention on alignment, work on disagreement in AI has also focused on divergence. For instance, large language models (LLMs) owe much of their success to the use of reinforcement learning with human feedback (RLHF), whereby the unsupervised output of these models is constrained by human evaluations of model outputs in the fine-tuning stage (Ouyang et al., 2022). Note, however, that this technique ultimately corresponds to divergence-reduction: The model is trained to prioritize outputs that agree with the humans providing feedback. This raises the worry that RLHF may "render models aligned 'on the surface', and that they still harbor harmful biases or other tendencies that may surface in more subtle contexts" (Bai et al., 2022). In other words, we could end up developing models that are radically misaligned, and undetectably so, because the models have been rendered incapable of expressing divergence from human judgment. In this way, divergence enables alignment and progress—an observation familiar to political scientists studying the 'spiral of silence' in the context of oppression (Noelle-Neumann, 1974). A key question for research on LLMs is therefore how models can be trained to express divergence—and hence enable misalignment detection—while maintaining usability. Work in the pragmatics of disagreement (Sifianou, 2012) and negotiations (Brett & Thompson, 2016) is highly relevant to making progress on this aim.

Artificial intelligence research is also uniquely positioned to answer questions about how data, divergence, and misalignment relate to each other, given the precision with which these metrics can be extracted from models. For instance, recent research indicates that many different neural networks trained on the same data not only seem to have high levels of agreement in their predictions, but also have similar "relative representations" (i.e. they exhibit high degrees of representational alignment), despite stochastic factors in their initialization and training processes (Moschella et al., 2023). Interestingly, this suggests that representational diversity in humans for common concepts (Marti et al., 2023) may arise more from diversity in our data (i.e., lived experience) than diversity in architectures (i.e., individual neural differences).

Recent work in AI has also explored disagreement and misalignment between humans and models, rather than across models. In many tasks, model performance is defined through convergence with human-generated labels (e.g., on categorization tasks). This work shows that supervised learning can increase model performance (i.e., convergence) while decreasing representational alignment with humans (Muttenthaler et al., 2022), though the relation between performance and alignment may be better characterized as a domain-dependent U-shaped function (Sucholutsky & Griffiths, 2023). Muttenthaler et al. (2022) also find that alignment is best predicted by the training dataset and objective function, whereas "model scale and architecture have essentially no effect," echoing the point above. Ultimately, these findings reinforce our previous observation that divergence and misalignment are correlated, but distinct dimensions of disagreement. Key divergences (e.g., whether the sun is at the center of the universe) can precipitate misalignment (e.g., about the structure of the solar system), but are neither necessary nor sufficient for misalignment, and vice versa.

Returning to the question of disagreement across diverse agents (e.g., humans and AI), these findings suggest the following conclusion: Divergence across agents is precipitated by misalignment, which is driven by differences in data, cognitive architectures, and goals. Modular collaboration thus poses an interesting challenge: If members of a team are exposed to highly differing data, perhaps because they are solving differing subgoals for the main task, they may develop differing representations, hindering communication. Thus, a promising area for future research is efficient policies for fostering alignment. For instance, datapoints that are highly informative in structuring the environment or that capture informative statistics of the space (e.g., prototypes) can be periodically shared across team members to anchor their representations. This entails a process of searching for representationally informative judgments, as well as sequencing these judgments in structured domains for optimal recovery, in a recipient-sensitive manner. Relatedly, team members can periodically be assigned to other goals to foster alignment.

## Conclusion

Disagreement arises from a complex interplay between divergence and misalignment, yet past research in judgment and decision making has largely focused on divergence. Recent work on misalignment in artificial intelligence has developed efficient methods for measuring and comparing representations across agents, and holds promise for enriching current research in judgment and decision-making and beyond. In particular, misalignment may play an important role in explaining biased advice-taking, the persistence of controversial beliefs, and algorithm aversion. Moreover, there are many unanswered questions about how we can infer and resolve misalignment in humans, artificial agents, and groups.

Galileo was forced to "abjure, curse, and detest" his dissent and sentenced to house arrest for the rest of his life (Finocchiaro, 2014), in part because the social structures of his time were designed to preserve stability, rather than promote progress. Developing a deeper understanding of disagreement can ultimately help us move beyond merely avoiding or suppressing divergence, and develop strategies for leveraging diverse perspectives toward solving difficult societal problems. Dissent, after all, is a sign of a healthy democracy, and a necessary precondition of productive diversity (e.g., Derex & Boyd, 2016).

# References

Atz, K., Grisoni, F., & Schneider, G. (2021). Geometric deep learning on molecular representations. *Nature Machine Intelligence*, 3(12), 1023-1032.

Bai, Y., Jones, A., Ndousse, K., Askell, A., Chen, A., DasSarma, N., ... & Kaplan, J. (2022). Training a helpful and harmless assistant with reinforcement learning from human feedback. arXiv preprint arXiv:2204.05862.

Bommasani, R., Hudson, D. A., Adeli, E., Altman, R., Arora, S., von Arx, S., ... & Liang, P. (2021). On the opportunities and risks of foundation models. arXiv preprint arXiv:2108.07258.

Bonaccio, S., & Dalal, R. S. (2006). Advice taking and decision-making: An integrative literature review, and implications for the organizational sciences. *Organizational behavior and human decision processes*, 101(2), 127-151.

Bovens, L., & Hartmann, S. (2003). Bayesian Epistemology. Oxford University Press.

Brett, J., & Thompson, L. (2016). Negotiation. Organizational Behavior and Human Decision Processes, 136, 68-79.

Carothers, T., & O'Donohue, A. (Eds.). (2019). *Democracies divided: The global challenge of political polarization*. Brookings Institution Press.

Carey, S. (1985). *Conceptual Change in Childhood*. MIT Press.

Carey, S. (1991). Knowledge acquisition: Enrichment or conceptual change. *The epigenesis of mind: Essays on biology and cognition*, 257-291.

Chi, M. T., Siler, S. A., & Jeong, H. (2004). Can tutors monitor students' understanding accurately?. *Cognition and instruction*, *22*(3), 363-387.

Derex, M., & Boyd, R. (2016). Partial connectivity increases cultural accumulation within groups. *Proceedings of the National Academy of Sciences*, *113*(11), 2982-2987.

Dietvorst, B. J., Simmons, J. P., & Massey, C. (2015). Algorithm aversion: people erroneously avoid algorithms after seeing them err. *Journal of Experimental Psychology: General*, 144(1), 114.

Duan, Y., & Lupyan, G. (2023). Divergence in Word Meanings and its Consequence for Communication. In Proceedings of the Annual Meeting of the Cognitive Science Society, 45.

Farrell, J., McConnell, K., & Brulle, R. (2019). Evidence-based strategies to combat scientific misinformation. *Nature climate change*, 9(3), 191-195.

Ferguson, J. (1756). *Astronomy Explained Upon Isaac Newton's Principles: And Made Easy to Those who Have Not Studied Mathematics*. T. Longman.

Frances, B., & Matheson, J. (2019)p. Disagreement. In E. N. Zalta (Ed.), The Stanford Encyclopedia of Philosophy. Metaphysics Research Lab, Stanford University.

Frith, C., & Frith, U. (2005). Theory of mind. *Current biology*, 15(17), R644-R645.

Gabriel, I. (2020). Artificial intelligence, values, and alignment. *Minds and machines*, 30(3), 411-437.

Gearhart, C. A. (1985). Epicycles, Eccentrics, and Ellipses: The Predictive Capabilities of Copernican Planetary Models. Archive for History of Exact Sciences, 32(3/4), 207–222. http://www.jstor.org/stable/41133751

Gershman, S. J. (2019). How to never be wrong. Psychonomic bulletin & review, 26, 13-28.

Glikson, E., & Woolley, A. W. (2020). Human trust in artificial intelligence: Review of empirical research. *Academy of Management Annals*, 14(2), 627-660.

Goldman, A. I. (2006). Simulating minds: The philosophy, psychology, and neuroscience of mindreading. Oxford University Press.

Grant, E. (2007). In defense of the Earth's centrality and immobility: Scholastic reaction to Copernicanism in the seventeenth century. American Philosophical Society.

Griffiths, T. L. (2020). Understanding human intelligence through human limitations. *Trends in Cognitive Sciences*, 24(11), 873-883.

Hartman, R., Blakey, W., Womick, J., Bail, C., Finkel, E. J., Han, H., ... & Gray, K. (2022). Interventions to reduce partisan animosity. *Nature human behaviour*, 6(9), 1194-1205.

Herrmann, E., Call, J., Hernández-Lloreda, M. V., Hare, B., & Tomasello, M. (2007). Humans have evolved specialized skills of social cognition: The cultural intelligence hypothesis. *Science*, 317(5843), 1360-1366.

Hütter, M., & Ache, F. (2016). Seeking advice: A sampling approach to advice taking. *Judgment and Decision Making*, 11(4), 401-415.

Iyengar, S., Lelkes, Y., Levendusky, M., Malhotra, N., & Westwood, S. J. (2019). The origins and consequences of affective polarization in the United States. *Annual review of political science*, 22, 129-146.

Jern, A., Chang, K. M. K., & Kemp, C. (2014). Belief polarization is not always irrational. *Psychological review*, 121(2), 206.

Kappes, A., Harvey, A. H., Lohrenz, T., Montague, P. R., & Sharot, T. (2020). Confirmation bias in the utilization of others' opinion strength. *Nature neuroscience*, 23(1), 130-137.

Kriegeskorte, N., Mur, M., & Bandettini, P. A. (2008). Representational similarity analysis-connecting the branches of systems neuroscience. *Frontiers in systems neuroscience*, 4.

Kuhn, T. S. (1962). *The structure of scientific revolutions*. University of Chicago Press.

Lake, B. M., Ullman, T. D., Tenenbaum, J. B., & Gershman, S. J. (2017). Building machines that learn and think like people. *Behavioral and brain sciences,* 40, e253.

Logg, J. M., Minson, J. A., & Moore, D. A. (2019). Algorithm appreciation: People prefer algorithmic to human judgment. *Organizational Behavior and Human Decision Processes*, 151, 90-103.

Marjieh, R., Van Rijn, P., Sucholutsky, I., Sumers, T., Lee, H., Griffiths, T. L., & Jacoby, N. (2022, September). Words are all you need? Language as an approximation for human similarity judgments. In The Eleventh International Conference on Learning Representations.

Mercier, H., & Sperber, D. (2011). Why do humans reason? Arguments for an argumentative theory. *Behavioral and brain sciences*, 34(2), 57-74.

Moschella, L., Maiorca, V., Fumero, M., Norelli, A., Locatello, F., & Rodola, E. (2022). Relative representations enable zero-shot latent space communication. arXiv preprint arXiv:2209.15430.

Muttenthaler, L., Dippel, J., Linhardt, L., Vandermeulen, R. A., & Kornblith, S. (2022). Human alignment of neural network representations. arXiv preprint arXiv:2211.01201.

Marti, L., Wu, S., Piantadosi, S. T., & Kidd, C. (2023). Latent diversity in human concepts. *Open Mind*, 7, 79-92.

Meißner, D. (2023). Plato's Cratylus. In E. N. Zalta (Ed.), The Stanford Encyclopedia of Philosophy. Metaphysics Research Lab, Stanford University.

Noelle-Neumann, E. (1974). The spiral of silence a theory of public opinion. *Journal of communication*, 24(2), 43-51.

Oktar, K. and Lombrozo, T. (2023). How Beliefs Persist Amid Controversy: The Paths to Persistence Model. [Manuscript submitted for publication].

Oktar, K., & Lombrozo, T. (2022). Mechanisms of Belief Persistence in the Face of Societal Disagreement. Proceedings of the Annual Meeting of the Cognitive Science Society, 44. Retrieved from https://escholarship.org/uc/item/3380n01h

Orchinik, R., Dubey, R., Gershman, S., Powell, D., & Bhui, R. (2023). Learning About Scientists from Climate Consensus Messaging. In Proceedings of the Annual Meeting of the Cognitive Science Society (Vol. 45, No. 45).

Ouyang, L., Wu, J., Jiang, X., Almeida, D., Wainwright, C., Mishkin, P., ... & Lowe, R. (2022). Training language models to follow instructions with human feedback. *Advances in Neural Information Processing Systems*, 35, 27730-27744.

Reeder, G. D., Pryor, J. B., Wohl, M. J., & Griswell, M. L. (2005). On attributing negative motives to others who disagree with our opinions. *Personality and Social Psychology Bulletin*, 31(11), 1498-1510.

Ross, L., & Anderson, C. A. (1982). Shortcomings in the attribution process: On the origins and maintenance of erroneous social assessments. In D. Kahneman, P. Slovic, & A. Tversky (Eds.), Judgment under uncertainty: Heuristics and biases (pp. 129 –152). Cambridge, England: Cambridge University Press.

Russell, S. (2019). Human compatible: Artificial intelligence and the problem of control. Penguin.

Sharma, A., Lin, I. W., Miner, A. S., Atkins, D. C., & Althoff, T. (2023). Human–AI collaboration enables more empathic conversations in text-based peer-to-peer mental health support. *Nature Machine Intelligence*, 5(1), 46-57.

Shepard, R. N. (1980). Multidimensional scaling, tree-fitting, and clustering. *Science*, 210(4468), 390-398.

Sifianou, M. (2012). Disagreements, face and politeness. *Journal of pragmatics*, 44(12), 1554-1564.

Simis, M. J., Madden, H., Cacciatore, M. A., & Yeo, S. K. (2016). The lure of rationality: Why does the deficit model persist in science communication?. *Public understanding of science*, 25(4), 400-414.

Sucholutsky, I., Battleday, R. M., Collins, K. M., Marjieh, R., Peterson, J., Singh, P., ... & Griffiths, T. L. (2023, July). On the informativeness of supervision signals. In Uncertainty in Artificial Intelligence (pp. 2036-2046). PMLR.

Sucholutsky, I., & Griffiths, T. L. (2023). Alignment with human representations supports robust few-shot learning. arXiv preprint arXiv:2301.11990.

Sutskever, I., & Leike, J. (2023, July). Introducing Superalignment. OpenAI Announcement. https://openai.com/blog/introducing-superalignment

Tenenbaum, J. B., Kemp, C., Griffiths, T. L., & Goodman, N. D. (2011). How to grow a mind: Statistics, structure, and abstraction. *Science*, 331(6022), 1279-1285.

Vosniadou, S., Vamvakoussi, X., & Skopeliti, I. (2008). The framework theory approach to the problem of conceptual change. *International handbook of research on conceptual change*, *1*, 3-34.

Yaniv, I., & Kleinberger, E. (2000). Advice taking in decision making: Egocentric discounting and reputation formation. *Organizational Behavior and Human Decision Processes*, 83(2), 260-281.